\documentstyle[11pt,newpasp,twoside,epsf]{article}
\def\edcomment#1{\iffalse\marginpar{\raggedright\sl#1\/}\else\relax\fi}
\reversemarginpar

\begin{document}
\title{The low-mass IMF  --  deep star counts in the dSph galaxy Ursa Minor}
 \author{Sofia Feltzing}
\affil{Lund Observatory, Box 43, SE-221 00 Lund, Sweden
            }
\author{Rosemary Wyse and Mark Houdashelt}
\affil{Physics \& Astronomy Department, Johns Hopkins University,
Baltimore, MD 21218, U.S.A.}
\author{Gerard Gilmore}
\affil{Institute of Astronomy, Madingley Road, CB3 0HA Cambridge, U.K.}

\begin{abstract}
 We present a new study of deep star counts in the Local Group dwarf
spheroidal (dSph) in Ursa Minor. Both the luminosity function (LF) and
the colour-magnitude diagram (CMD) of the unevolved stars are compared
with the LF and CMD of the old, metal-poor globular cluster M92. The
main sequence locations and turn-offs are identical within the
errors. Since we know from the brighter evolved stars  that the
metallicities for  these two disparate systems are the same this
implies that they also have equal ages.  A direct comparison of faint
LFs is then equivalent to comparison of the low-mass stellar Initial
Mass Functions (IMF).  We find that their LFs are identical within the
mass-range covered ($\sim 0.35 - 0.8~M_\odot$). The Ursa Minor dSph
has one of the highest apparent M/L ratios known in the Local Group,
and is an extremely low surface brightness external galaxy. M92 is a
typical high surface brightness globular cluster, with no apparent
dark matter. These results lead to the conclusion that the low-mass
stellar IMF in systems that formed at high redshift is independent of
environment.  Indeed, it is consistent with the low-mass IMF in
star-forming regions today.
\end{abstract}

\section{Introduction}

In the Local Group of galaxies some 20 dwarf spheroidal galaxies
(dSph) have been identified (Mateo, 1998). They are some of the most
uninteresting objects, to the eye, in the sky with their low surface
brightness (around 25 magnitudes per square arcsec in V) spread over a
large angle (up to 160 arcmin for Sextans). Yet, they may hold answers
to the quest to understand the nature of dark matter. As reviewed
recently by Mateo (1998) Local Group dSph galaxies have some of the
largest known M/L ratios for  a single galaxy, and an order of
magnitude larger than that derived for  normal spiral  galaxies. These
large M/L ratios have been inferred from the observed velocity
dispersion in the central parts of these dSph galaxies and are all in
excess of the value expected for  purely stellar systems
with a standard IMF.

The questions then naturally arise of how do these high M/L ratios
arise,  are they due to truly high dark matter contents or are there
other  explanations possible? Several alternative  suggestions have
been put forward; e.g.  ``artificially'' inflated velocity dispersions
by tidal effects and projection effects (e.g. Klessen \& Kroupa 1998)
or by unresolved binary stars, though a very non-standard binary
population is required (Hargreaves et al. 1996).  True dark matter in
these small systems must be cold,  but standard non-baryonic CDM
provides for dark haloes that are too centrally-concentrated
(e.g. Moore 1999).  Baryonic dark matter remains a possibility, and
here we will investigate whether or not the apparent dark matter
simply can be in the form of low-mass stellar objects.  The method we
have chosen for our investigation is simple; obtain a deep luminosity
function of a dSph with lots of apparent dark matter and compare that
to the luminosity function of a system known to contain no dark
matter, such as a globular cluster.

The dSph chosen should have as narrow as possible ranges of age and
 metallicity as a mixture of several ages and metallicities would make
 the experiment more complicated and parameter dependent.  Further, we
 want the dSph to have large amounts of apparent dark matter and the
 Ursa Minor and Draco dSphs are the two outstanding examples with
 $(M/L)_0=$ 60 and 58 $(M_{\odot}/L_{\odot})$ as measured 
in the V-band, respectively (Mateo,
 1998,  Table 4).   Detailed spectroscopic abundance analysis has been
 done for a few stars in both Draco and Ursa Minor (Stetson 1984 and
 Shetrone et al. 1998, 2001).  The Ursa Minor dSph has, within the errors,
 one metallicity, [Fe/H]$=-2.2\pm0.2$ dex (Stetson 1984),  as compared
 to Draco which has a clearly detected spread in metallicity of about
 one dex. Stetson (1984) finds a spread of $=-2.9<$[Fe/H]$<-1.8$ and
 the recent Keck observations by Shetrone et al. (1998, 2001) confirms this
 result.  Furthermore the Ursa Minor dSph  has a simple star-formation
 history (consistent with one single early burst of star formation)
 and no second parameter problem (see e.g. Mateo, 1998) and a
 negligible reddening, $E(B-V)\simeq 0.03$ (Zinn, 1981). It is also
 close, $(m-M)=19.1\pm 0.1$  (Olszewski \& Aaronson, 1985) which means
 that it  is possible to reach deep down on the main sequence.

\section{The data}

Observations were obtained using the WFPC2 on board HST during 1997,
1998, and 1999 in the two filters F606W and F814W (HST program
GO~7419, PI~Wyse).  The images were processed through the standard
pipeline.  The final, drizzled images have total exposure times in
F606W of 14600 sec and in F814W of 17200 sec. These data supersede the
earlier, partial dataset presented in Feltzing et al. (1999).
Photometry was derived from the drizzled images using the IRAF DAOPHOT
package. Empirical {\sl psf}s were created individually for each
filter and image and the $\chi$ and sharpness parameters were used to
eliminate e.g. unresolved background galaxies.  The {\sl
psf}-photometry in this way provided a coordinate list that was then
used to derive new {\sl psf} fitted photometry using the scheme
described by Cool \& King (1996). The corresponding colour-magnitude
diagrams  were then constructed by cross-correlating the stellar
coordinates. A total of 2038, 1751, and 1698 stars were detected in
this way on WF2, WF3, and WF4, respectively.  Completeness has been
estimated by adding artificial stars to the drizzled images and then
rerunning the  detection and photometry procedure described above on
the artificial images.  The magnitudes for the retrieved artificial
stars were compared with  their input magnitudes and only stars with
input $-$ output $=$ 0.5 magnitudes were kept when completeness was
determined.

\begin{figure}
\plottwo{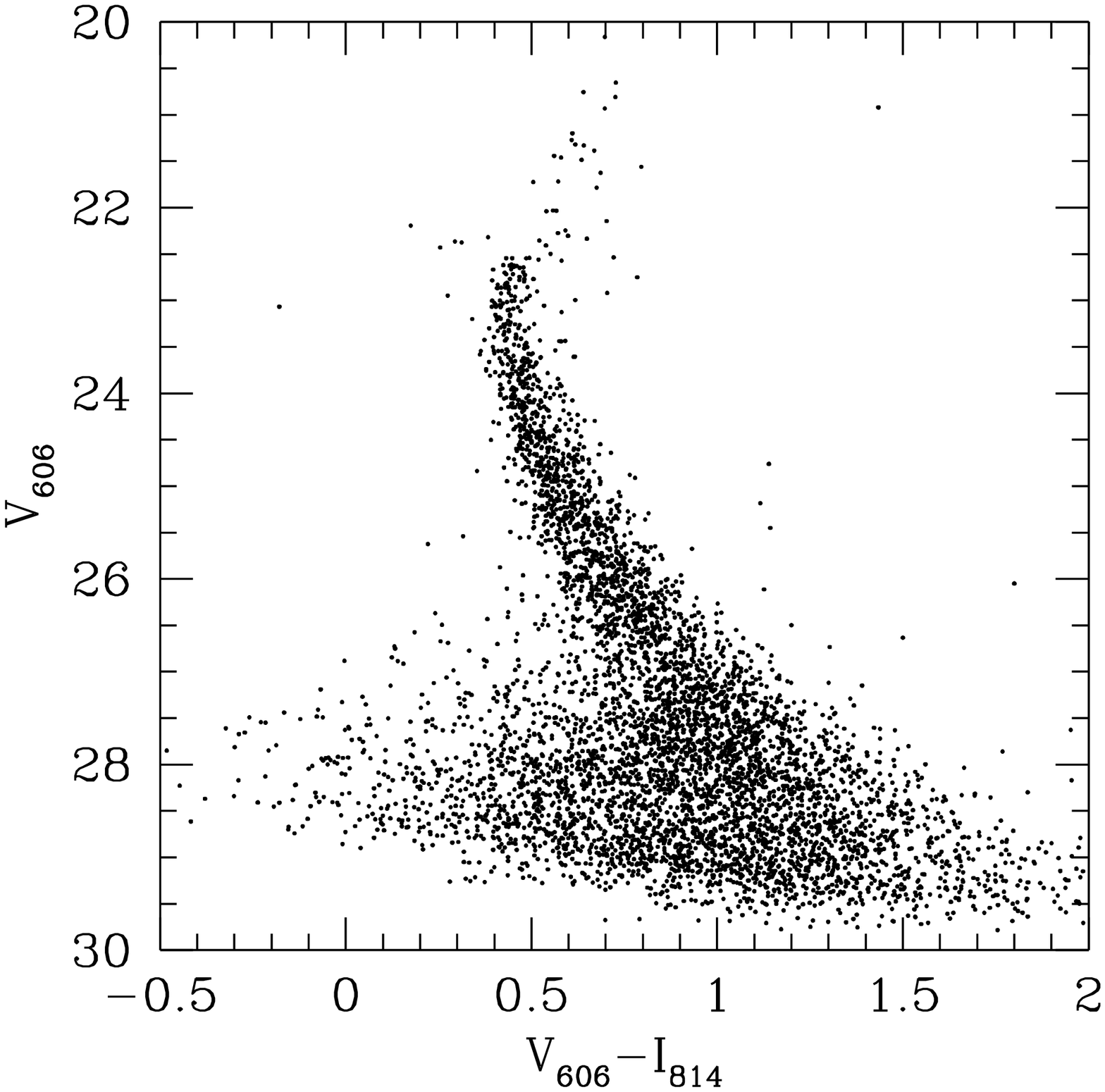}{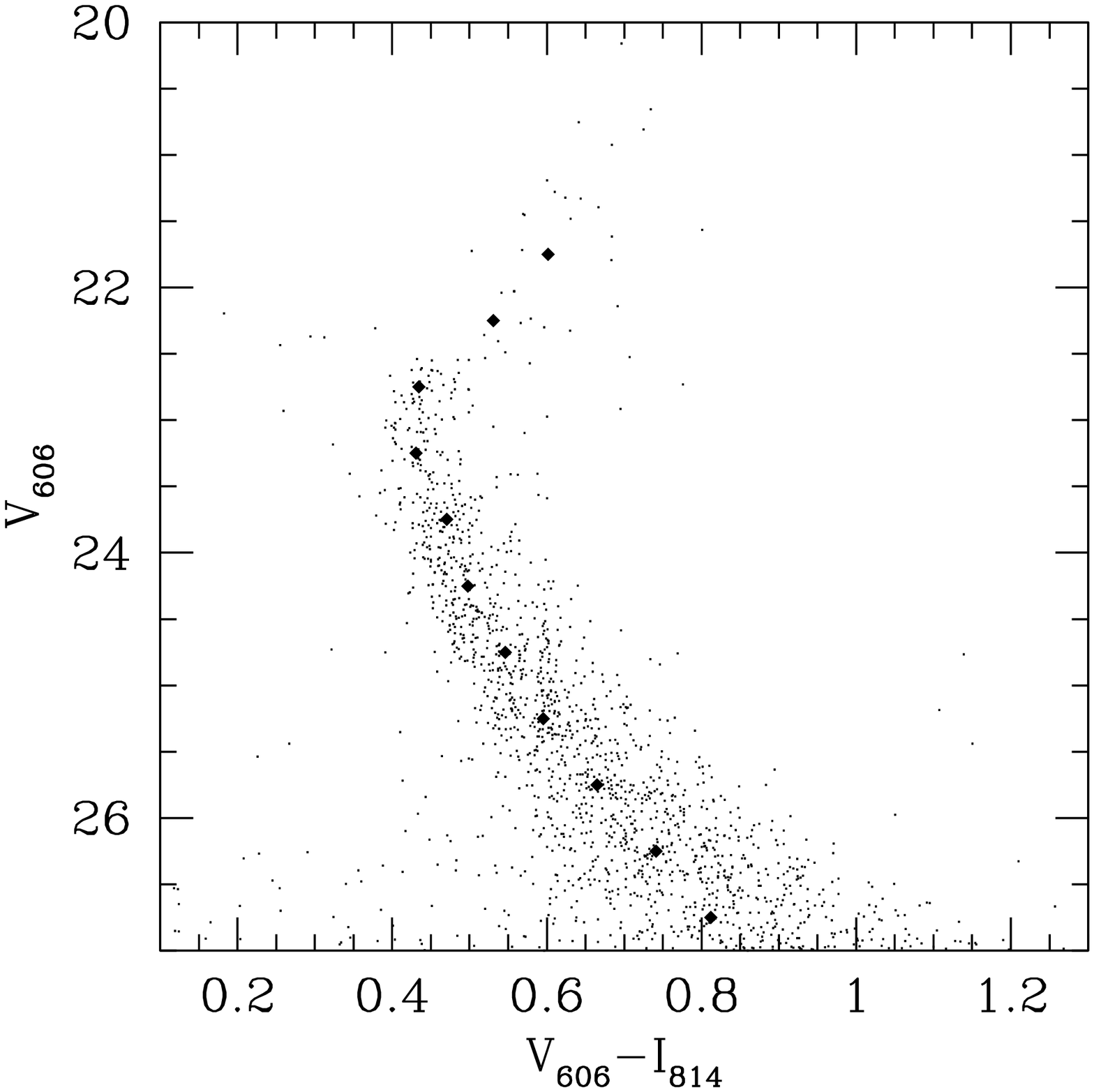}
\caption{CMD for the Ursa Minor dSph. In the right-hand panel 
we show a blow-up 
including a comparison with the fiducial line from the M92 CMD,
based on data from Piotto et al. (1997).}
\end{figure}

\begin{figure}
\plotfiddle{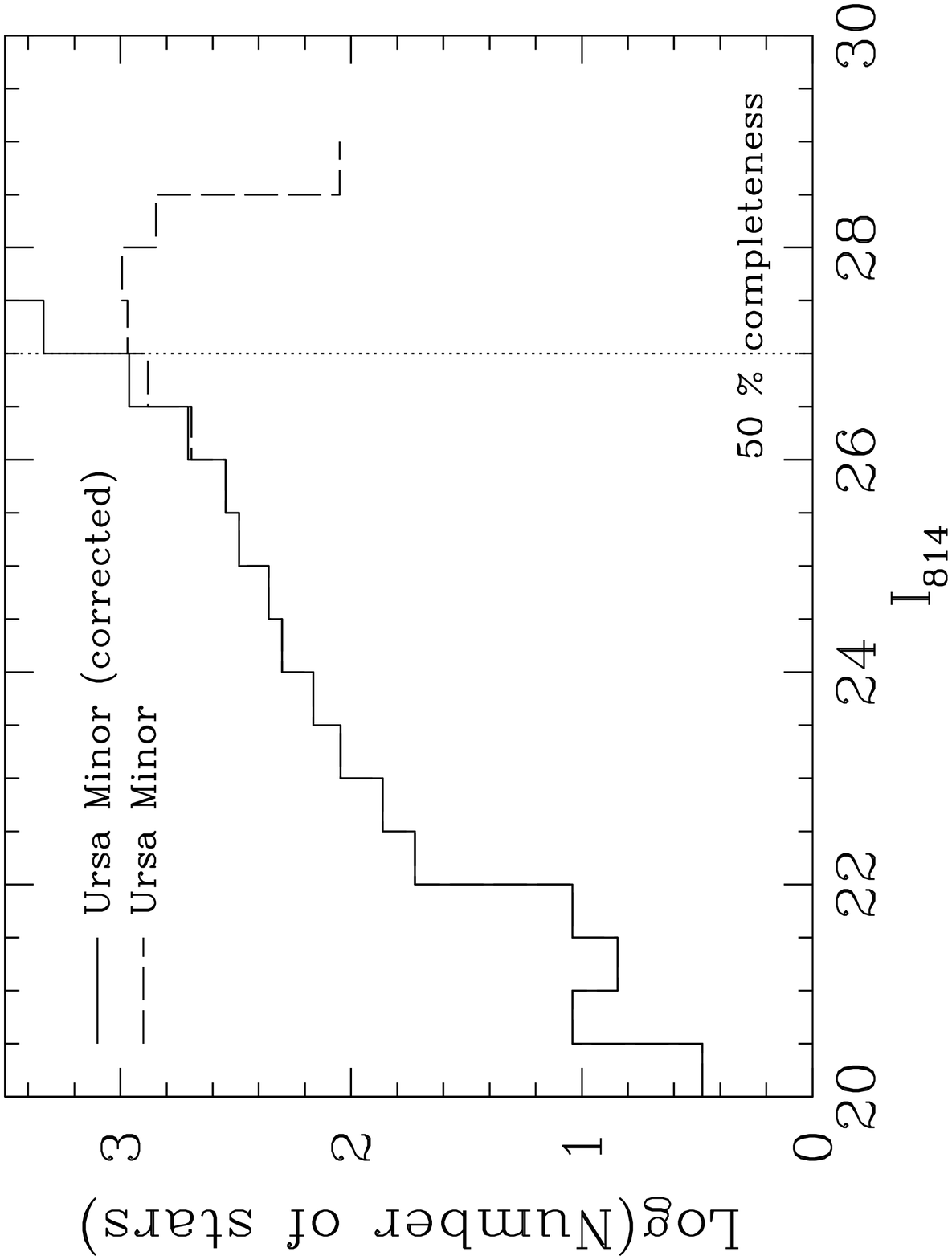}{2cm}{-90}{22}{22}{-170}{60}
\plotfiddle{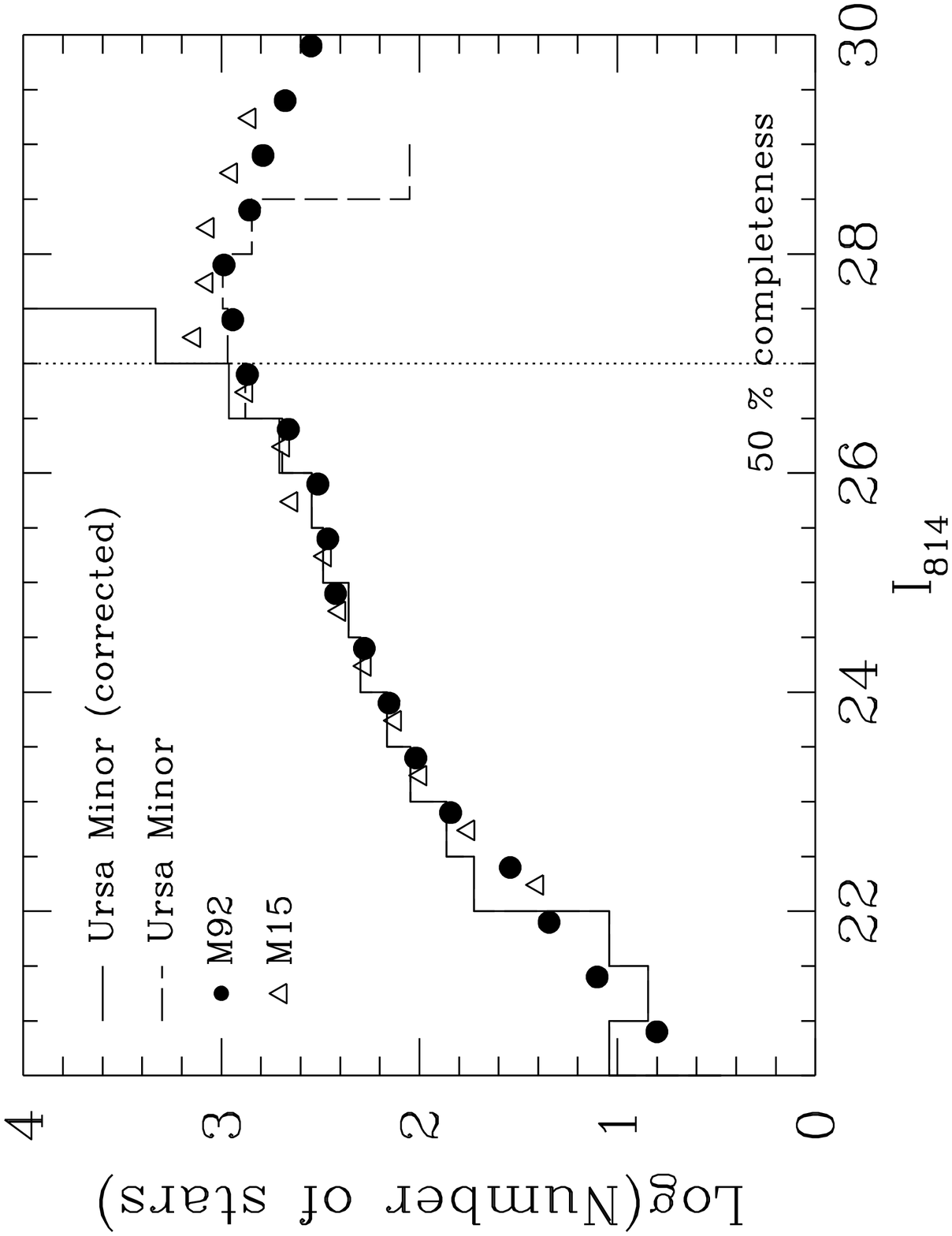}{2cm}{-90}{22}{22}{00}{131}
\caption{Left panel shows the uncorrected and corrected luminosity
function for the Ursa Minor dSph in the I band. The right-hand
panel shows the same luminosity functions as well as the LF from
the globular cluster M92 as well as that from M15 (Piotto et al. 1997).}
\end{figure}

The luminosity function for each WF was then constructed from the
CMDs. The data were, for each WF,  then corrected for incompleteness
and then  all three LFs were combined into one, shown in Fig. 2.  The
final comparisons between M92 and the Ursa Minor dSph is shown in Fig.
1b (CMD)  and 2b (LF).  The M92 data have been moved to the same
distance modulus as that of the Ursa Minor dSph and renormalized.  It
is clear from this that down to the 50\% completeness limit there is
no significant difference between the number of stars at a given
magnitude (and thus inferred mass) in the two systems (Fig. 2b). This
means that in a system with a high amount of inferred dark matter,
Ursa Minor, and in a system with no dark matter, M92, the stellar
formation processes are such that the relative numbers of low mass
stars are  the same. Furthermore, these two systems -- a globular
cluster and a dSph galaxy -- are at opposite extremes of stellar
number density, with the central V-band surface brightness of M92
being 15.6 mag/sq arcsec (Harris 1996), while that of the Ursa Minor
dSph is 25.5 mag/sq arcsec (Mateo 1998).

As the metallicity of both the Ursa Minor dSph and M92 are known and
are very similar we may therefore compare their colour magnitude
diagrams to determine whether or not they have differing ages,
Fig. 1b. Since the two objects have the same metallicity this plot
infers that they have the same age as well. This confirms the recent
results by Mighell \& Burke (1999).

\section{Binaries}

\begin{figure}
\plottwo{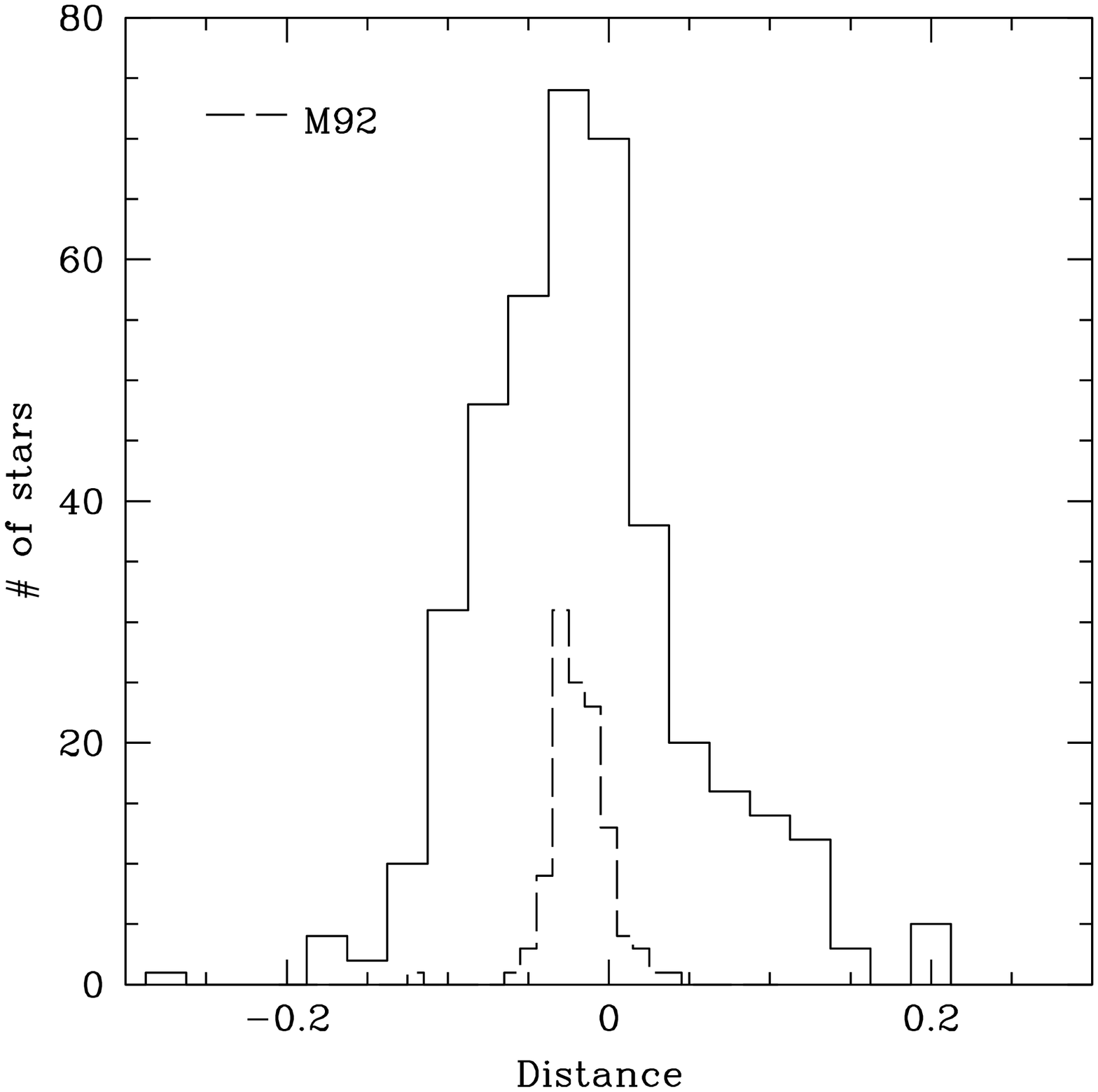}{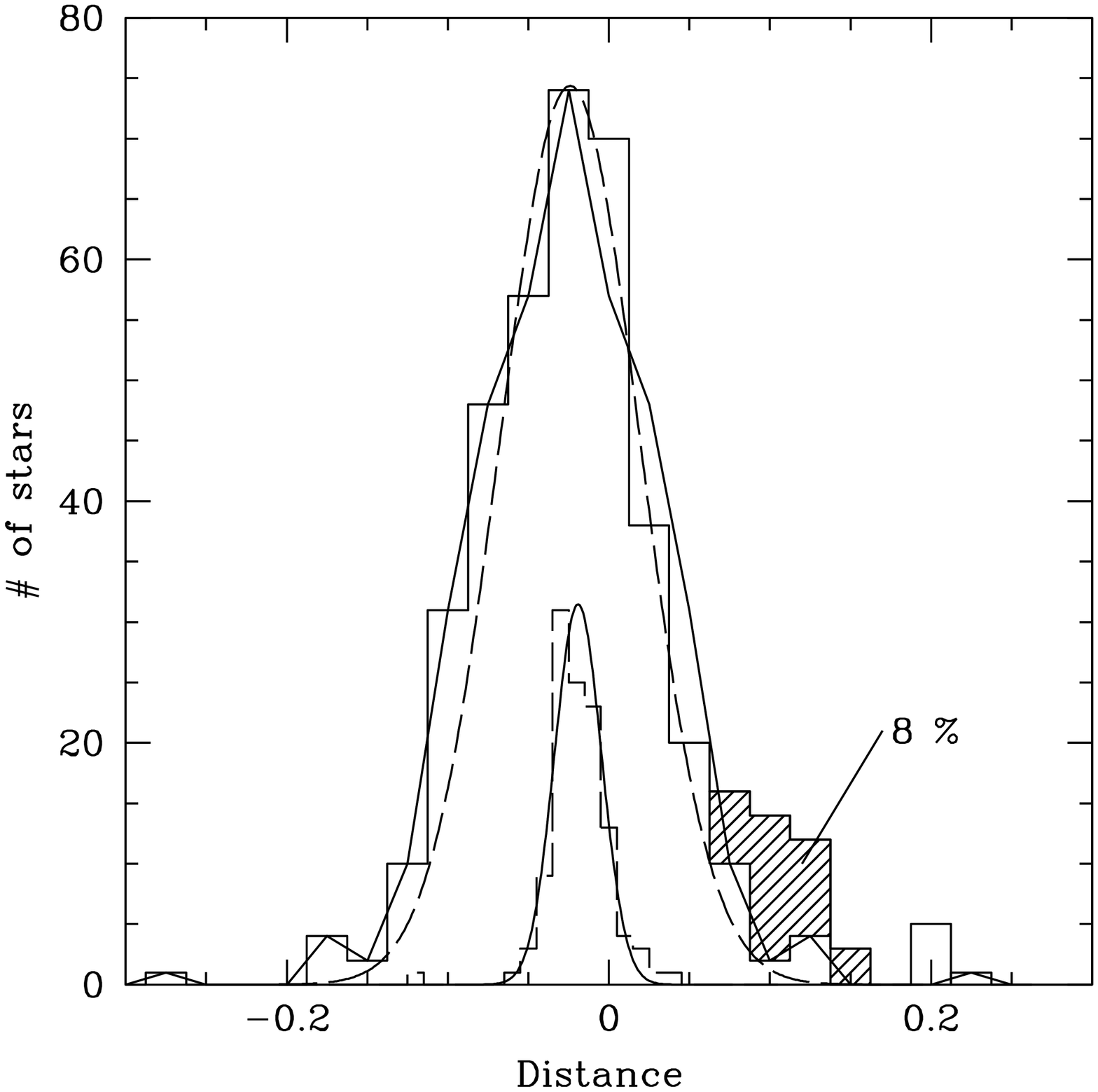}
\caption{Left panel shows the histogram of the stars perpendicular 
distances of each stars from the fiducial ridge line of the main sequence
for both the Ursa Minor dSph and the globular cluster M92. 
The right-hand panel shows the same
diagrams but also the same Gaussian fitted to the two histograms (full line
for M92 and dashed for the  Ursa Minor dSph)
as well as fit to the blue side of the Ursa Minor histogram that
then has been mirrored on to the red side (full line).
Finally the binary stars are marked as a shaded area. }
\end{figure}

If binary systems are present in a stellar population they may be
distinguished in the CMD. The easiest case to detect would arise if
all binary systems were made up of two stars with equal mass. In this
case the binaries would make up a sequence in the CMD that would have
the same colour as the individual stars in the system  but $2.5 \log
(2)$ magnitudes brighter than the main sequence. If the systems are
made up of stars of unequal masses there will be a gradual spread of
stars between the main sequence and the equal mass sequence Hurley
\& Tout  (1998).  Although $V-I$ is not the best discriminatory
between single and binary stars (because the main sequence becomes
quite vertical in this colour) a spread to the red can clearly be seen
in our colour magnitude diagram.

We now proceed to quantify this visual impression. First we determined
a  fiducial ridge line by binning in $V$ and finding the mean
$V-I$ for each bin. This was done using the data for the full CMD
for all three chips. Then we selected a clean portion of the main
sequence, $24.8 \la V \la 25.8$. In this magnitude range we 
do not have to concern ourselves with completeness issues.
For each star in the selected part
of the CMD and for each chip we then calculated the distance to the
ridge line. Then we determined the median distance from the ridge line
on each chip. Minor differences were found, such that $median(WF4)
-median(WF2)=0.010 $ and $median(WF4) -median(WF3)=-0.026 $. The
median for chip four is $-0.016$. That the median is  negative is to be
expected as the fiducial ridge line has constructed using all stars,
both binary and single, and is thus redder  than the actual main
sequence.

A histogram of the distances was then
constructed, Fig. 3. The histogram shows a rather typical
distribution for a single stellar population with binaries present. It
has a fairly steep, Gaussian looking blue side, reaching a flat-ish
peak (binaries of unequal mass),  then falling off fairly rapidly and
showing a tail consisting of equal mass binaries. That these features
are correctly interpreted is illustrated by the small histogram for
the M92 data that is also shown. M92 is  known to present at least an
8\% binary candidate fraction (Romani \& Weinberg, 1991) as derived
from ground based   CCD photometry.

To estimate the number of equal mass binary systems in the Ursa Minor dSph we
simply mirrored the blue part of histogram to the red side and counted
the number of stars in the ``bump'' on the red side,
Fig. 3. In this way we find that 8 \% of the
systems observed are equal mass (or close to equal mass) binaries.
These are shown as the shaded area in Fig. 3.

Thus it appears that the Ursa Minor dSph has a normal binary 
population, indicating that undetected binaries are not the source of the 
large observed velocity dispersion. 

\section{The initial mass function in the Ursa Minor dSph}

The question to be answered in this study was whether or not the
initial mass function (IMF) in the Ursa Minor dSph was different from
that in systems so far studied. While our data was not expected to
reach to the brown dwarf limit, the indications from the solar
neighbourhood (Reid et al. 1999) is that there is a smooth continuity
across the M-star -- brown dwarf regime, so that should very low mass
objects be associated with the dark matter in this dSph, one might
expect to see a signal even in the M-dwarf regime to which our data
are sensitive.  The comparison with M92 shows that there is indeed no
excess of low mass stars  in the Ursa Minor dSph, down to $\sim 0.35
M_{\odot}$ (using isochrones from Baraffe et al. 1997).

We also quantify this by studying our complementary STIS (see Wyse et
al. 1999; 2001 in prep.) observations  and comparing them to a
theoretical mass function, Fig. 4.  A mass function of slope $-1.35$
(where Salpeter is $-2.35$) is  converted into an LF using the Baraffe
et al. (1997) models, Fig. 4.  The fit to the M92 and hence the Ursa
Minor dSph data is good. This  is in agreement with what was found for M92
and M15 by Piotto \& Zoccali (1999). Further, Zoccali et al. (2000)
found this to be an excellent fit to their data for the  faint LF of
the galactic Bulge, further supporting the universality of the IMF
of old stars,
across two to three dex in metallicity.  This slope is close to that of 
the solar neighbourhood IMF derived by Scalo (1986), over this mass range.

\begin{figure}
\plotfiddle{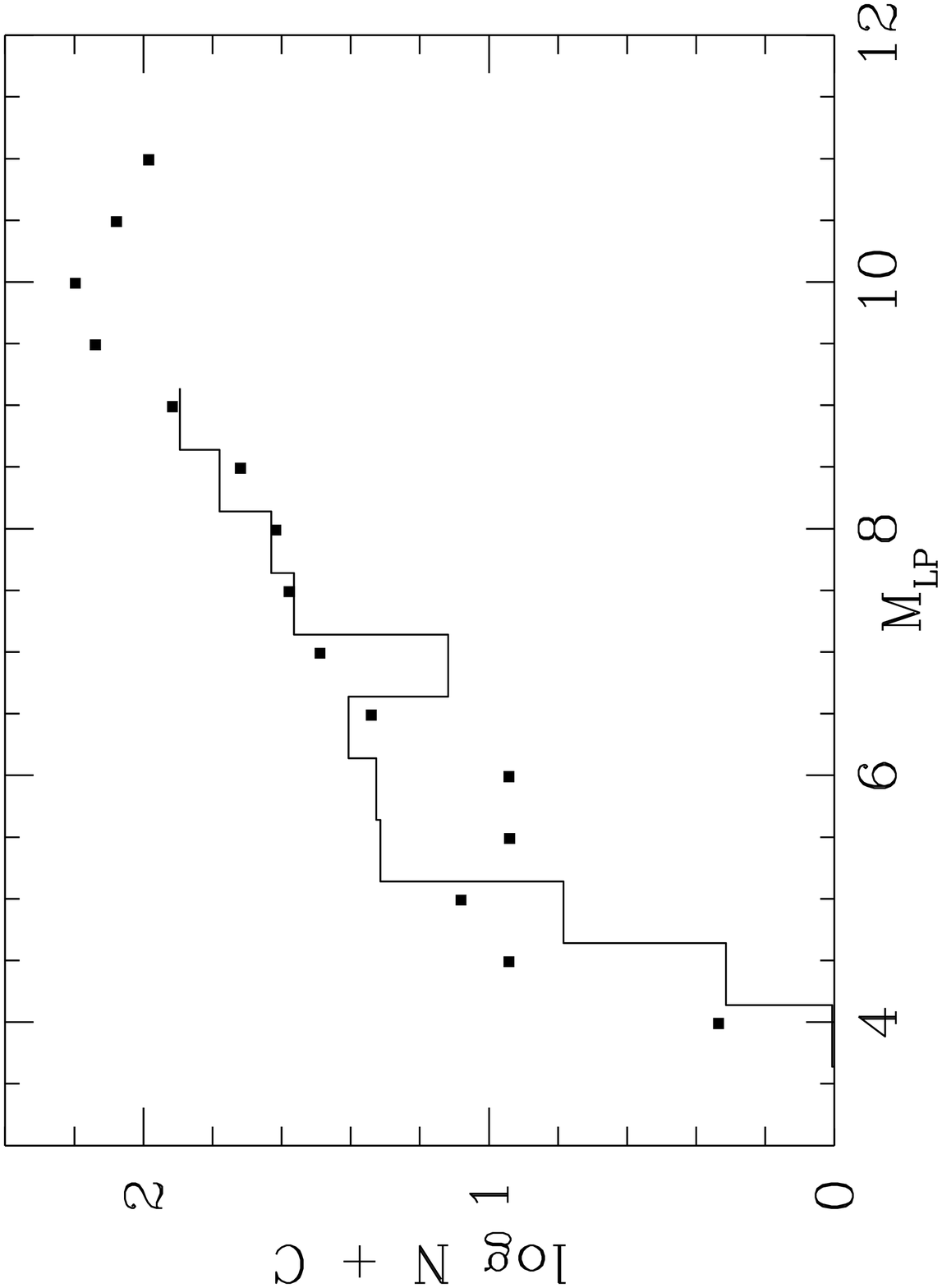}{2cm}{-90}{25}{25}{-190}{70}
\plotfiddle{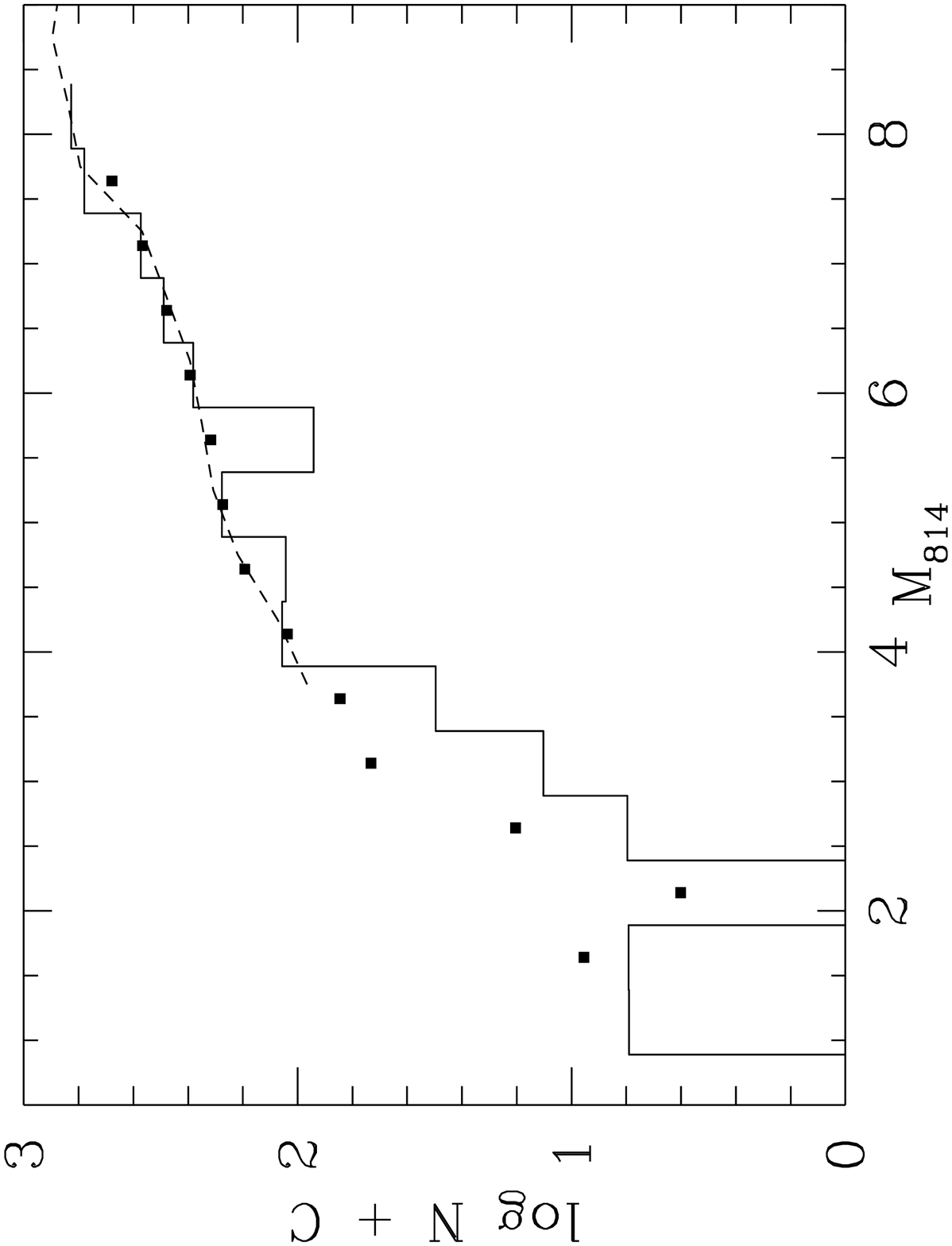}{2cm}{-90}{25}{25}{00}{141}
\caption{Left panel shows the comparison between the completeness
corrected STIS LP LF for the Ursa Minor dSph (solid line) and the globular 
cluster M15 (filled squares). The right-hand panel shows the 
a comparison of the STIS-based LF for the Ursa Minor dSph (solid line=
with the LF from the WFPC2 data (filled squares). Both are then
compared with a theoretical LF that corresponds to a power-law 
mass function with a slope of $-1.35$ (dotted line). }
\end{figure}

\section{Conclusions}

We have done a purely differential study between the faint stellar LF
of  the  Ursa Minor dSph  and that of the  metal-poor and old globular
cluster M92. In summary we find that (a) the luminosity function in
the Ursa Minor dSph is the same, at low masses, as that in the
globular cluster M92 which leads to the conclusion  that the apparent
dark matter in the Ursa Minor dSph can not be made up of low mass
stars (b)  that the Ursa Minor dSph has the same age as globular
cluster  M92, and (c) that there are binaries present in the Ursa
Minor dSph, and that equal mass binaries make up $\sim 8$ \%.

Since this LF and underlying IMF are remarkably similar to that of 
e.g. the galactic bulge these results intriguingly appear to further 
indicate the universality of the IMF. In particular it is similar over
several dex in metallicity as well as in environments of very different
densities, e.g. a tenuous dSph and a dense globular cluster.

\noindent
{\bf Acknowledgement}
Based on observations with the NASA/ESA Hubble Space Telescope, obtained at 
STScI, operated by AURA Inc, under NASA contract NAS5-26555. 
Support for this work was provided by NASA grant
number GO-7419 from STScI.

\end{document}